\documentclass{article}

\usepackage{microtype}
\usepackage{graphicx}
\usepackage{subfigure}
\usepackage{subcaption}
\usepackage{xurl}
\usepackage{booktabs} 

\usepackage{hyperref}

\usepackage[accepted]{icml2024}

\usepackage{amsmath}
\usepackage{amssymb}
\usepackage{mathtools}
\usepackage{amsthm}

\usepackage[capitalize,noabbrev]{cleveref}

\theoremstyle{plain}

\theoremstyle{definition}

\theoremstyle{remark}

\icmltitlerunning{Science of Data Collection}

\begin{document}

\twocolumn[
\icmltitle{Position: Insights from Survey Methodology can Improve Training Data}



\icmlsetsymbol{equal}{*}

\begin{icmlauthorlist}
\icmlauthor{Stephanie Eckman}{a1}
\icmlauthor{Barbara Plank}{a2,a3,a4}
\icmlauthor{Frauke Kreuter}{a5,a4,a1,a6}
\end{icmlauthorlist}

\icmlaffiliation{a1}{Social Data Science Center, University of Maryland, College Park, MD, USA}
\icmlaffiliation{a2}{Center for Information and Language Processing (CIS), LMU Munich, Germany}
\icmlaffiliation{a3}{Computer Science Department, IT University of Copenhagen, Denmark}
\icmlaffiliation{a4}{Munich Center for Machine Learning (MCML), LMU Munich, Germany}
\icmlaffiliation{a5}{Institute for Statistics, LMU Munich, Germany}
\icmlaffiliation{a6}{Joint Program in Survey Methodology, University of Maryland, College Park, MD, USA}

\icmlcorrespondingauthor{Stephanie Eckman}{steph@umd.edu}

\icmlkeywords{Training Data, Data Quality, Surveys, Noisy Labels, Annotation}

\vskip 0.3in
]



\printAffiliationsAndNotice{}  

\begin{abstract}
Whether future AI models are fair, trustworthy, and aligned with the public's interests rests in part on our ability to collect accurate data about what we want the models to do. However, collecting high\--quality data is difficult, and few AI/ML researchers are trained in data collection methods. Recent research in data\--centric AI has show that higher quality training data leads to better performing models, making this the right moment to introduce AI/ML researchers to the field of survey methodology, the science of data collection. We summarize insights from the survey methodology literature and discuss how they can improve the quality of training and feedback data. We also suggest collaborative research ideas into how biases in data collection can be mitigated, making models more accurate and human\--centric.

\end{abstract}

\section{Introduction}
Social scientists have long relied on survey data collected from human subjects to quantify the population, understand public opinion, and test hypotheses about human behavior. The methods used to collect survey data have been extensively studied and refined by researchers in the field of \emph{survey methodology}, which draws on social and cognitive psychology to develop theories about how humans understand, process, and respond to questions in surveys \citep{Groves2009-co}. 

Data labeled by humans is also central to all stages of the AI pipeline \citep{plank-2022-problem, mazumder2023dataperf}, from initial model training, to fine-tuning, reinforcement learning, and model assessment. Insights from social science can contribute to the development of more trustworthy and human\--centric models: ``if we want to train AI to do what humans want, we need to study humans'' \citep{Irving2019}. However, collecting high\--quality data is difficult, as decades of research in survey methodology and recent high-profile failures in opinion polling \citep{BritishPolls, USPolls2016, USPolls2020} demonstrate. 

Given the importance of human\--labeled data to AI model development, we are surprised that little research in the AI literature has used social science, and survey methodology in particular, to understand the actions and motivations of the humans behind the data generating process. We worry that many researchers collecting data to train, fine-tune, or reinforce AI and ML models are not trained in data collection. A recent paper lamented that, among AI researchers, ``everyone wants to do the model work, not the data work'' \citep{data_work}. 

\textbf{This position paper argues that lessons from survey methodology can improve the quality and efficiency of training data and thus improve models trained on those data.} We introduce AI researchers to the community of scientists who want to do the data work and their insights into how to collect high\--quality data. We first make the case that label collection is similar to survey data collection (Section \ref{sec:likesurvey}). Next, we draw on social science theories to develop hypotheses about the facets of the data collection task that may impact the quality of the labels collected (Section \ref{sec:measurement}). Then, we discuss who works as labelers and how the characteristics and uniqueness of the labelers can impact the labels collected and the models trained on those data (Section \ref{sec:representation}). In Section \ref{sec:transparency}, we join the call for greater transparency in label collection methods, offering lessons from surveys and statistical methods. Throughout the paper, we use the terms labels (and labelers) to refer generally to ML annotations, such as image object labels and bounding boxes, natural language understanding labels, model evaluation data, human feedback for reinforcement learning, and other types of training data (and data generators).

\section{How Labeling is Like a Survey}\label{sec:likesurvey}

\begin{figure*}[htp]
    \centering
    \includegraphics[width=1\linewidth]{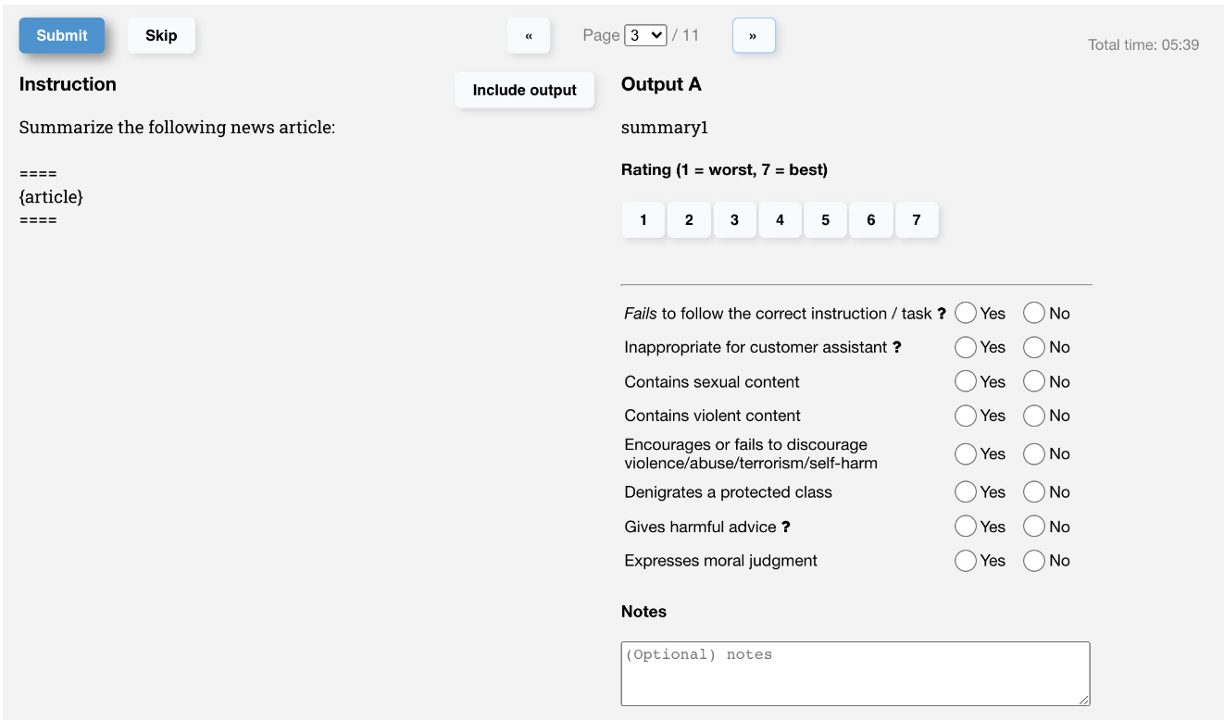}
    \caption{Example of Labeling Interface for InstructGPT \citep[from][]{ouyang2022training}}
    \label{fig:instructgpt}
\end{figure*}

Surveys can be self-administered, like a web or paper\--and\--pencil survey, or interviewer administered, like a telephone or face-to-face survey. A web survey presents a series of questions with, most often, closed answer choices. 
Superficially, the process of labeling observations for ML models often looks like a web survey: labelers see one or more prompts and associated answer choices (Figure \ref{fig:instructgpt}). As in surveys, the task can be factual (is there a bicycle in this photo?) or opinion based (which of these responses to the prompt is most helpful?). 


Of course, label collection differs from survey data collection in important ways. Surveys ask many different questions and end when the questions run out; labeling tasks usually consist of similar repeated observations, and labelers often continue until they choose to stop. Surveys usually ask people about themselves: their opinions, behaviors, characteristics. Labeling tasks more often involve passing judgment on an object outside of oneself (for example, images, product reviews, or a news article). However, the human feedback data used in reinforcement learning often aims to capture personal opinions: ``which of the following responses is most relevant to the prompt?'' 

The goal of surveys and labeling tasks is also different. Surveys ask questions of a sample of selected persons to make inference about the population. For example, say we wished to know what proportion of the U.S. adult population does not have health insurance. We select a sample of U.S. adults and ask them whether they have health insurance. For each respondent, we have a yes or no response to the question.\footnote{We are ignoring ``I don't know'' and other types of responses.} The proportion of respondents without insurance in the sample is an estimate of the proportion of the population without insurance. A survey often asks many questions on related topics. Our example survey might also ask ``when was the last time you visited a health care professional?'' and ``have you received a flu vaccine in the last 12 months?'' Results are often reported at the question level although relationships between questions are also of interest: for example, are people with health insurance more likely to get a flu vaccine? The quality of the analysis rests on collecting \textit{accurate responses} from a \textit{representative sample}.

The goal when collecting labeled data is not to estimate the proportion of the population that finds a given post offensive or detects a vehicle in a given image. Instead, the goal is to learn patterns from the labeled data to predict labels for unseen observations. Thus, while it is not as important that labels reflect the population's views at the observation level, across observations, the views of the population should be represented. Collecting \textit{accurate labels} from a \textit{diverse set of labelers} is important to the performance and generalizability of the final model.

Despite their differences, both surveys and label collection have these two needs in common. The data points provided in response to a prompt should capture the data provider's judgment: accurate responses / accurate labels. Those who provide the data points should represent the judgments of the relevant population: representative sample / diverse set of labelers. 
Sections \ref{sec:measurement} and \ref{sec:representation} discuss these needs in turn.

\section{Need for Accuracy}\label{sec:measurement}

Survey methodologists have developed theories of how respondents understand, process, and respond to survey questions. We summarize these theories and use them to derive hypotheses about the aspects of the label collection task that may impact label quality. We end the section with thoughts on mitigation measures and future research. 

\subsection{Response Process}\label{sec:theories}

Ideally, survey respondents understand questions thoroughly and respond thoughtfully. However, they make take shortcuts that can threaten data quality.

\paragraph{Optimal Survey Response Process}
Responding to a survey question can involve several cognitive steps \citep{tourangeau2000psychology, Tourangeau2018}:
\begin{enumerate}
    \item Comprehension: Understand the question and the response options
    \item Retrieval: Search memory for relevant information
    \item Integration: Integrate the retrieved information to form an answer to the question
    \item Mapping: Map that answer onto the provided answer choices 
\end{enumerate}
Ideally, a respondent proceeds through each step in order. However, they can choose to backtrack. For example, considering the response options in the Mapping step may change the interpretation of the question (Comprehension) or trigger additional relevant information to come to mind (Retrieval). 

The above model exposes why respondents sometimes give incorrect answers. At the Comprehension step, they may fail to understand the question or some of the words it uses. They may have a different understanding of some of the words than those who wrote the question. At the second step, respondents may fail to retrieve all relevant information. Some information may have been forgotten. At the third step, respondents may fail to put in the mental effort to bring together their understanding of the question with the retrieved information. At the fourth step, respondents may not find an answer choice that reflects their answer or they may edit the true answer to avoid revealing sensitive information. 

\paragraph{Deviations from Optimal Response Process}
The full survey response process outlined above is cognitively demanding. Some respondents resort to taking shortcuts, an approach called satisficing \citep{Krosnick91, Krosnick1996}. For example, they may retrieve only the most recent relevant information from memory (recency bias) or choose the first reasonably correct answer choice. Satisficing relates to the cognitive miser theory in psychology, which holds that people seek to minimize cognitive effort \citep{Fiske1991-rj, kahneman2011thinking}. 

As predicted by satisficing theory, eye\--tracking studies show that respondents do not read all options in \textit{select\--all\--that\--apply} questions \citep{Galesicetal2008}, and shortcuts are more common as survey length increases \citep{Galesicetal2009}. Respondents tend not to read provided instructions \citep{Brosnanetal19} or click on provided definitions \citep{Peytchevetal2010}, especially when they believe they understand the concept that is asked about \citep{Touragenauetal2006}.

The survey literature discusses several types of more extreme undesirable response behavior. Acquiescence is the tendency to say ``yes'' to \textit{yes/no} questions, regardless of content~\citep{Knowles_acq}. Straightlining is the practice of choosing the same response option in the same position (for example, the first response option) to all questions. This behavior is most common in batteries or grids of questions with the same response options \citep{Kim2019}. Some respondents even deliberately give incorrect answers to later questions to reduce the length or burden of a survey: when a ``yes'' response triggers follow up questions, respondents may learn to report ``no.''  This phenomenon is called motivated misreporting \citep{KMPT, Tourangeau2012, Eckman2014}. 

\paragraph{Context Effects}
Perceptions and judgments are shaped by the broader context and preceding experiences, a phenomenon called context effects \citep{tversky1974judgment, strack1992}. 
For example, a very tall person can make others seem shorter: a contrast effect. An unethical politician can make other politicians seem less ethical: an assimilation effect \citep{blessschwarz2010}. 

Opinion questions are especially vulnerable to context effects, because respondents do not always have well\--formed, fixed opinions that they retrieve from memory. Instead, they form opinions when asked for them, and this process can be shaped by context clues in the question, the response options, the look and feel of the instrument, or the previous questions \citep{McFarland81, Zaller1992, Schwarz2007}.

Order effects are the most common example of context effects in surveys. Questions that come earlier in a survey can change how respondents interpret later questions. Researchers have found order effects in reports of crime and bullying victimizations, of disabilities, and of race \citep{Cowanetal78, Gibsonetal78, Batesetal95, Todorov00, Huang_order}.

\subsection{Hypotheses about Label Quality}\label{sec:hypotheses}
These theories about how respondents answer questions lead us to several hypotheses about the properties of the labeling task that may impact training data quality. The ML literature has investigated some of these hypotheses, but fundamental research gaps exist. 

\paragraph{Wording and Reading Level} 
Respondents cannot provide high\--quality answers to questions unless they clearly understand what the question asks and what they should include and exclude in their answer. Questions should be at an eighth grade reading level or lower \citep{Dillmanetal2014} and terms should be as unambiguous as possible. Definitions, if needed, should be provided in the question itself, because respondents often do not use rollovers or links for additional information when answering questions \citep{Peytchevetal2010}. 

We suspect that applying the same guidelines to labeler prompts and instructions would improve the quality of the labels collected. (Of course, nothing can be done about ambiguous terms or high reading levels in the observations.) We are not aware of any research into these issues. We note that Figure \ref{fig:instructgpt} contains rollovers or links for additional information on three of the questions on the right side. 

\paragraph{Multiple Labels} 

\begin{figure}[ht!]
\subfigure[Collect all labels on one screen]{\includegraphics[width=3in]{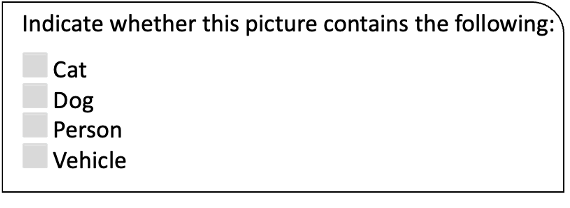}\label{fig:1screen}}
\subfigure[Collect labels on separate screens]{\includegraphics[width=3in]{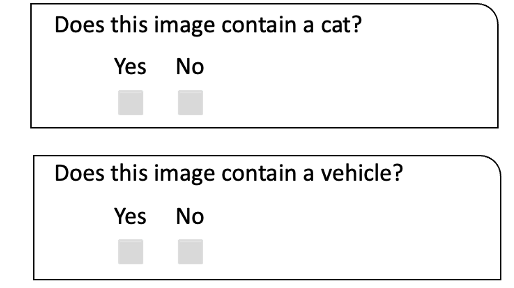}\label{fig:sep_screens}}
\caption{Collecting multiple labels on one screen (first panel) or multiple (second panel); adapted from \citep{kern-etal-2023-annotation}}
\label{fig:multi}
\end{figure}

Often ML researchers want to collect multiple labels about an observation from one labeler: for example, whether an image contains a cat, a dog, a person, or a vehicle. We can ask labelers to provide all labels at once, as in Figure \ref{fig:1screen}, or we can ask one or more labelers to provide each label separately, as in Figure \ref{fig:sep_screens}. 

\begin{figure}[ht!]
\subfigure[\textit{Select\--all\--that\--apply} question]{\includegraphics[width=3in]{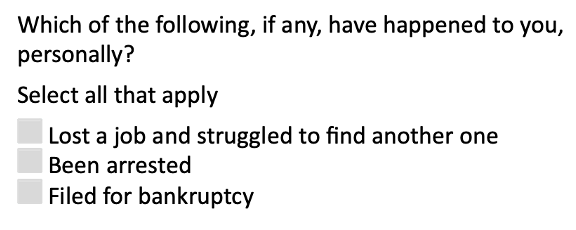}\label{fig:select_all}}
\subfigure[Series of \textit{yes/no} questions]{\includegraphics[width=3in]{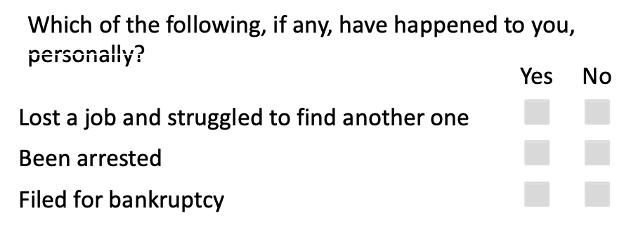}\label{fig:yesno}}
\caption{Survey question in \textit{select all} (first panel) and \textit{yes/no} (second panel) formats, adapted from \citet{pew_selectall}}
 \label{fig:selectall}
\end{figure}

The choice between these two approaches echos the choice in surveys between \textit{select\--all\--that\--apply} questions (Figure \ref{fig:select_all}) and a series of \textit{yes/no} questions (Figure \ref{fig:yesno}). As predicted by the survey response model, the \textit{yes/no} format collects better data because it encourages respondents to process each option separately \citep{Smyth2006, pew_selectall}. The \textit{select\--all\--that\--apply} approach is vulnerable to satisficing: respondents pick the first one or two reasonable options and fail to think deeply or even look at later options \citep{Galesicetal2008}.
However, the \textit{yes/no} format can encourage acquiescence \citep{Smyth2006}.

This finding also holds for labeling. In an experiment that involved labeling tweets as containing hate speech or offensive language, \citet{kern-etal-2023-annotation} randomly assigned labelers to different versions of the labeling instrument. Condition A was similar to Figure \ref{fig:1screen} and Condition B was similar to Figure \ref{fig:sep_screens}. Splitting the collection across two screens (Condition B) led to higher rates of hate speech and offensive language annotation. Models trained on Condition B data also performed better than those trained on Condition A data across several metrics \citep{kern-etal-2023-annotation}. This result is a clear example of how findings in the survey literature translate to the labeling task and improve the quality of training data.

\paragraph{Order Effects} 
Theories about context effects suggest that the order in which instances or observations are presented influences the labels assigned. If a contrast effect is present, a very hateful social media post would make later posts seem less hateful than they otherwise would \citep[for preliminary evidence of this phenomenon, see][]{beck2022}. An order effect could also arise if labelers change their behavior over time. As they gain experience, they might become more accurate and conscientious, as suggested by \citet{lee-etal-2022-annotation}. Alternatively, they might become bored or fatigued and engage in more satisficing, acquiescence, or ``don't know'' nonresponding over time, as suggested by the survey literature \citep{Kraut1975, Galesicetal2009}.

\citet{mathur-etal-2017-sequence} detected order effects in two benchmark NLP data sets. When coding tweets as hate speech or offensive language, \citet{beck-etal-2024-order} found a decreasing time trend: tweets that appeared later were less likely to be flagged. However, their experimental set up did not allow them to test hypotheses about the mechanisms causing the time trend.

Many research questions thus remain open, such as when contrast and assimilation effects appear and which tasks and labelers are most impacted by order effects. Order effects may also have implications for active learning (AL) and similar labeling approaches. In AL, an algorithm determines which observations to label next to maximize the marginal information gain for the model \citep{monarch_book}. However, active learning considers only the model's needs, not the labelers'. If context effects are large, the algorithm should also account for contrast and assimilation effects when deciding which observations to label. We are not aware of any research that has jointly accounted for the needs of the model for diverse training data and the impact of observation order on annotators. An approach similar to active learning exists in surveys \citep{Zhangetal20}, but is not widely adopted due in part to concerns about order effects.

\paragraph{Don't Know Option}
The inclusion of ``don’t know'' or ``no opinion'' responses in surveys has been debated for years. Some researchers believe these options offer respondents an easy way to satisfice: rather than thinking about the issue and forming an opinion, respondents can simply choose the ``don't know'' option. Others believe that having no opinion on a given topic is a valid response and that forcing respondents to provide an opinion when they don't have one reduces data quality \citep{Schuman96}. 

Many labeling tasks do not include a ``skip'' or ``don't know'' option: labelers must provide a label even when they are not certain. (The instrument in Figure \ref{fig:instructgpt} is an exception.) When a recent experiment provided a ``don't know'' option to half of the labelers, fewer than three percent chose it, and the overall distribution of the labels was not impacted \citep{beck2022}. Another recent study in NLP collected ``uncertain'' flags from labelers for a relation extraction task across several text genres~\cite{bassignana-plank-2022-crossre}. Labelers were more likely to choose ``uncertain'' when coding text in some genres, and the model struggled with prediction in those genres as well. These preliminary results suggest that giving labelers the option to indicate uncertainty or lack of knowledge can provide helpful information and does not encourage satisficing in labeling.

\paragraph{Pre-labeling}
Pre-labeling involves displaying a suggested label, bounding box, or similar and asking the labeler if it is correct. If the labeler indicates the label is not correct, they are asked to provide the correct label. Pre-labeling is more efficient than labeling without suggested labels \citep{Lingren2013, Skeppstedt2016, SOUTH2014}. However, labelers may become too trusting of the suggestions and fail to correct errors \citep{Dietvorst2015, logg2017, berzak-etal-2016-anchoring}, a phenomenon called anchoring bias or automation bias \citep{Mosier1999}.

In the survey field, we find that providing a pre-filled response that respondents or interviewers should update leads to underreporting of errors of both omission and commission. For example, when respondents are reminded of their answer in a previous survey wave, they tend to report that the answer still applies rather than providing an updated response \citep{Jckle2019}. 

Previous literature on labeling has explored anchoring bias \citep{Lingren2013, Skeppstedt2016, SOUTH2014} but has not leveraged social science to find the factors that make the effect weaker or stronger. The social science literature suggests several hypotheses about the mechanisms behind anchoring bias, such as incentives \citep{cialdini2009influence}, belief in authority \citep{asch2016effects, cialdini2009influence}, or reliance on heuristics \citep{cialdini2009influence, Norman2007, kahneman2011thinking}. Testing these theories experimentally would help data collectors design tasks that capture the efficiency of pre-labeling with lower risk of anchoring bias.

\paragraph{Overreliance on Examples} 
Examples can introduce a similar bias. Survey questions often give examples of the things respondents should consider when they formulate their responses. Examples improve response accuracy when they remind respondents to include items they might otherwise leave out, because they have forgotten or were unsure whether to include them. However, when the examples include only common items, respondents tend to leave out less common items \citep{Tourangeauetal14}.

Examples are also often included in labeling instructions or annotation guidelines. As in surveys, labelers at times rely too heavily on these examples as they label. This \emph{instruction bias} can lead to overestimation of model performance \citep{parmar-etal-2023-dont}. Again, a better understanding of the mechanisms behind this behavior, guided by social science theories, could inform efforts to reduce it. 




\subsection{Mitigation Measures and Future Research}\label{sec:ideas1}
The survey methods literature suggests several approaches to minimize the effects discussed in Section~\ref{sec:hypotheses}.

\paragraph{Randomization of Observations} To address order effects, label collectors can randomize the order of observation shown to labelers. This approach does not eliminate order effect but it ensures that no one ordering impacts all annotators in the same way. Random ordering is incompatible with active learning techniques, however.

\paragraph{Instrument Testing} Many surveys spend weeks or even months drafting, testing, and revising questions and response options to arrive at language that is understood similarly by most members of the population, a process called cognitive interviewing \citep{Willis2004}. They then launch the survey with a small group of respondents to assess response rates, don't know rates, and response times. Such testing could improve the instructions and prompts given to labelers. 

\paragraph{Retain Paradata} Many surveys capture process data, called paradata, during the survey, such as the time spent on each screen, the device used, even mouse movements \citep{Kreuter2013, HorwitzKreuterEtAl2017}. Paradata can help identify satisficing respondents and low quality data \citep{Kreuter2013} and may do the same in label collection. However, collecting such data may raise additional privacy and ethics concerns \citep[see][]{Couper_Singer_2012, KunzetalParadata, Henningeretal23}.

\paragraph{Feedback to Labelers} Label collection instruments could experiment with prompts to encourage labelers not to engage in satisficing. Respondents who pick many ``don't know'' answers or repeatedly choose the same response option could receive reminders about the importance of the task. Those who click through screens  quickly could receive prompts to slow down and read carefully. In surveys, feedback on speeding successfully slowed respondents and did not lead to early terminations of the survey \citep{ConradetalSpeed}.



\paragraph{Test Observations} In surveys, instructed response items, such as ``Choose yellow below'' can help identify respondents who speed or provide low quality responses \citep{Gummer2018, berinsky_measuring_2024}. We have not seen  these questions used in labeling tasks. However, some tasks embed observations with known labels to try to catch annotators who do not understand the task. In NLP tasks, it is common to qualify only workers who pass an initial quiz or perform well on inserted test observations~\cite{nangia-etal-2021-ingredients}. These test observations could also catch annotators who satisfice \citep[see][for an application]{nie2020learn}.

We recommend future research to more comprehensively test these and related approaches.

\section{Need for Diversity}\label{sec:representation}

Large-scale annotation tasks, such as the reCAPTCHA tests, may collect labels from a broad spectrum of the population. However, the crowdworkers used in many label collection tasks are members of large crowdworker panels such as Appen, Upwork, Scale AI, Prolific, or MTurk and do not reflect the U.S. or world population. \citet{smart2024discipline} note that labelers tend to be from the Global South, while the models they help train benefit the educated population in the Global North. The workers who labeled data to fine-tune InstructGPT were 22\% Filipino and 22\% Bangladeshi \citep[][Appendix B3]{ouyang2022training}. MTurk members are younger, lower income and less likely to live in the South than the U.S. population \citep{Berinsky_Huber_Lenz_2012}. 

The unique characteristics of the labelers lead us to worry that the data they provide may not represent the views of the population that will use or be affected by the models.\footnote{We acknowledge that this section glosses over what we mean by ``population.'' Is it the population that regularly uses the models? The population impacted by the models? We leave this important discussion to later work.} Issues of representativeness are enormously important to surveys, which explicitly aim to make statements about the entire population. For this reason, the survey methods literature has much to contribute on this topic.

\subsection{Selection Bias}
Selection bias occurs when those involved in providing data have different characteristics than the population. In surveys, selection bias\footnote{In the survey literature, the preferred terms is nonresponse bias. We use the term selection bias here because it is more general and more suited to the labeling task.} arises when the propensity to take part in a survey is correlated with the characteristics measured in a survey. Let us return to the example in Section \ref{sec:likesurvey}, a survey to estimate the proportion of U.S. adults without health insurance. If we distributed the survey invitations in doctors' waiting rooms, the sample proportion would overestimate the population proportion. Those in waiting rooms are more likely to receive the survey invitation and are also more likely to have health insurance. The propensity to take part is correlated with what the survey measures, leading to selection bias in the estimate of health insurance coverage. 

\begin{figure}
\centering
\includegraphics[width=3in]{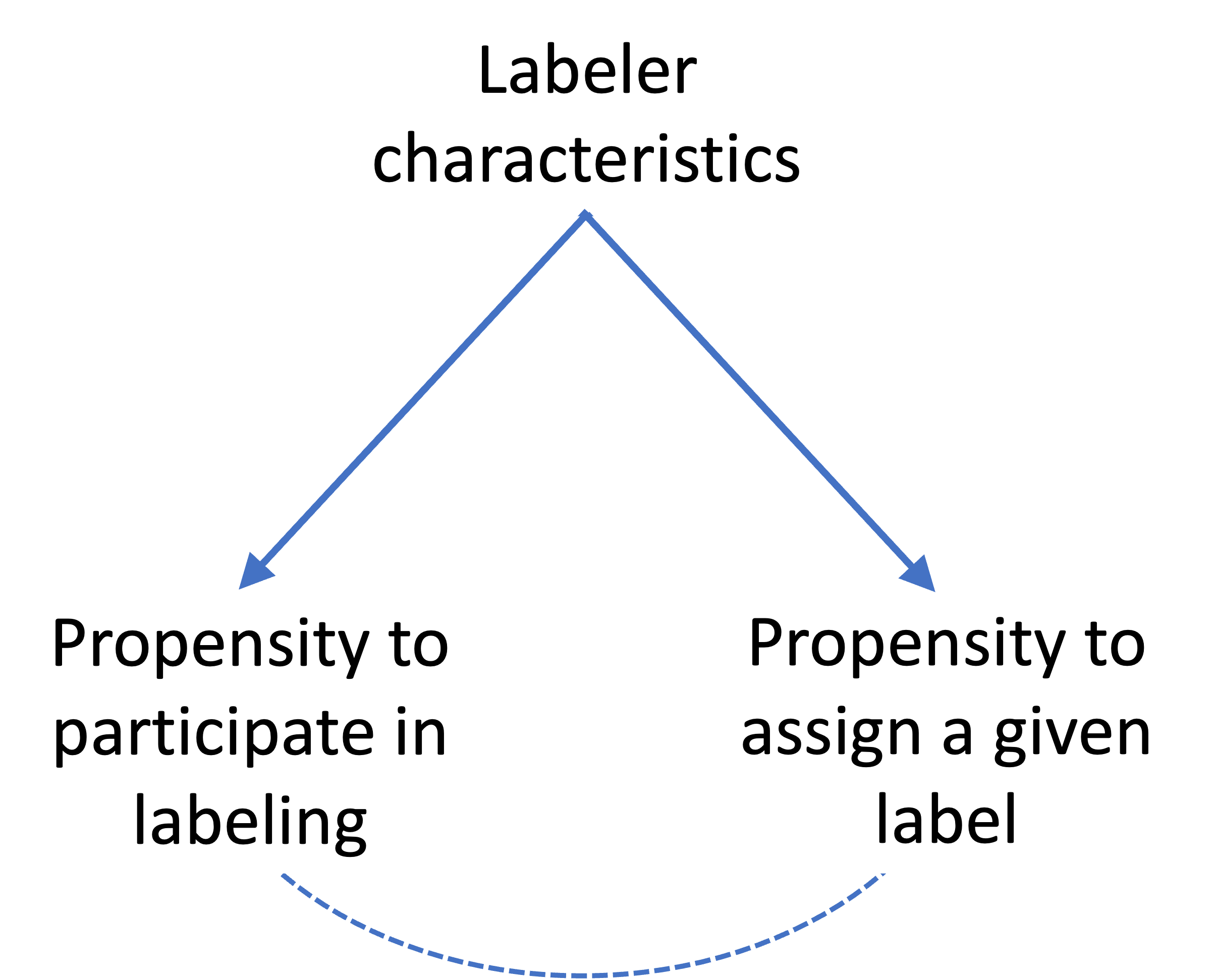}
\caption{Labeler characteristics induce correlation between propensities \citep[adapted from][]{Groves2006}}
\label{fig:correlation}
\end{figure}

Just as in surveys, selection bias arises in training data labels if the propensity to engage in the labeling task is correlated with the propensity to assign a given label. We expect that this correlation is non-zero for many tasks, because labeler characteristics likely influence both propensities, as shown in Figure \ref{fig:correlation}. 


On the left side of Figure \ref{fig:correlation}, labelers' characteristics influence their decision to engage in labeling work. As noted above, labelers tend to have different attributes than the general population \citep{gray_ghost_2019, smart2024discipline}. On the right side of Figure \ref{fig:correlation}, labelers' characteristics influence the labels they assign. The literature provides evidence for this association as well: labeler age and education level impact whether they perceive comments on Wikipedia entries as attacks \citep{kuwatlyetal}; conservative labelers are less likely to flag anti-Black language \citep{sap-etal-2022-annotators}; labelers in the U.S. are more likely to see a bird in ambiguous images than those in India~\cite{parrish-etal-2024-picture}.
We hypothesize that such effects are more widespread than the literature suggests, because many studies do not collect annotator characteristics and thus cannot detect their impact on the labels \citep{kirk2023personalisation}.

If the characteristics that influence one side of Figure~\ref{fig:correlation} are the same as, or correlate with, the characteristics that influence the other side of the figure, the two propensities at the bottom will be correlated, leading to selection bias. Consider a labeler who is a frequent biker, annoyed by cars that park in the bike lanes in their city. They may be more likely to agree to label a data set of potential bike lane violations (left side) and also more likely to see violations where others do not (right side).


Although the goal of label data collection is not to make population estimates, selection bias is nevertheless a risk to training data and model development. In the early days of machine learning, when developers trained model to recognize written numbers or tell cats from dogs, perhaps labeler characteristics mattered less. The association on the right side of Figure \ref{fig:correlation} may be weak or absent with more objective tasks \citep[though see][for counterarguments]{Aroyo_2015}.

However, the labeling tasks still performed by humans today often involve more difficult and more opinion\--based work. People may legitimately disagree about whether a given statement is toxic or offensive, for example. Reinforcement learning with human feedback (RLHF) in particular may be more exposed to selection bias. The labels collected for reinforcement learning, like those in Figure \ref{fig:instructgpt}, are inherently opinion-based. As discussed in Section 2, opinion questions in surveys are more susceptible to context effects than factual questions. We suspect that opinion labels are also more impacted by labeler characteristics than are more objective labels. RLHF aligns models to the judgments of labelers. If selection bias is present in the data, those judgments do not match the interests of the public. For example, if those who participate in labeling are also less likely to judge text as toxic, then the model trained on their data will also see less toxicity. 

We see evidence of the impact of selection bias on models in two studies that trained models on different sets of labelers. Both showed that models make different predictions when trained on labels from, for example, female versus male labelers, or Asian versus white labelers \citep{kuwatlyetal, Perikleous2022}.
 

\subsection{Mitigation Measures and Future Research}\label{sec:ideas2}
To combat selection bias in labeling tasks, we need to break the correlation between the propensity to assign a given label and the propensity to participate in labeling. We consider three  methods.

\paragraph{Left Side} We could try to remove the correlation between labeler characteristics and the propensity to participate in labeling (the left side of Figure \ref{fig:correlation}) by diversifying the labeler pool, collecting data from labelers with different motivations and characteristics. This approach is central to surveys: we solicit responses from a random sample of the population through appeals to public service, tokens of appreciation, and multiple reminders \citep[see, for example,][]{Groves1998}. If we can collect responses from a random sample of population members, selection bias is not a problem. Unfortunately, it is not clear that this approach will work with label collection: many people are not interested in labeling data for AI models. Surveys also find it increasingly difficult to recruit representative samples of respondents \citep{De_Leeuw_E2018-gy, Williams2018-at}.

\paragraph{Right Side} Another way to reduce selection bias is to break the correlation on the right side of Figure \ref{fig:correlation}, which would mean removing (or, more reasonably, reducing) the influence of labeler characteristics on the labels that they assign. More diverse examples in the instructions, use of test observations, feedback to labelers, and training in implicit bias might help labelers label more uniformly. (The literature on coding in qualitative studies takes a similar approach \citep[see, for example,][]{Hak1996}.) Even if these interventions do work to reduce the impact of labelers' characteristics on the labels they provide, however, they are expensive and do not scale well.

Interestingly, reducing selection bias by removing the correlation on the right side of the figure is not of interest in surveys, which aim to capture the diversity of respondents' behavior, opinions, and judgments. Recent research in NLP has similarly found that capturing the diversity of labels across labelers can improve models~\citep{basile2021we, sap-etal-2022-annotators}. \citet{Aroyo_2015} and \citet{plank-2022-problem} have argued that such human label variation is in fact \textit{information} and can improve model performance and trustworthiness.

\paragraph{Weighting} A third method to address selection bias is statistical adjustment. If we condition on the labeler characteristics in Figure \ref{fig:correlation}, we can remove (or reduce) the induced correlation between the two propensities. Like the right side approach, this method involves embracing labeler subjectivity and uses weights to get the balance of those characteristics right. 


The survey literature contains many statistical methods to match the characteristics of the respondents to the population and thus reduce selection bias \citep[][for example]{Bethlehem2011}. For example, surveys in many countries collect more responses from women than men; we use weights to ensure that the contribution of women's and men's responses on the final estimate matches their shares in the population. Future work could test the usefulness of these weighting approaches for improving machine learning models.\footnote{The tendency for some crowdworkers to give false answers to demographic questions to protect their privacy, reported by \citet{huang-etal-2023-incorporating}, will complicate any weighting approach. Misreporting of demographics also causes problems in surveys \citep{pew_demos}.} 

However, these statistical adjustments can work only if we capture the labeler characteristics that drive the relationships in Figure \ref{fig:correlation}. Thus we need a better understanding of what motivates people to work as labelers and what types of tasks are vulnerable to selection bias, which points to another role for social science to play in improving label quality.

\section{Transparency in Label Collection}\label{sec:transparency}

The discussion above suggests that labels are sensitive to how studies design the labeling task and recruit labelers, in ways often not recognized in the AI/ML literature. For this reason, we call for more transparency and documentation in how labels are collected when new data sets or models are released. 

The survey industry in the U.S. has embraced transparency in recent years. The American Association for Public Opinion Research launched the Transparency Initiative in 2014. Member firms agree to disclose details about how survey data were collected, such as question and response option wording and order, respondent recruitment protocols, and weighting adjustments.\footnote{See \url{https://aapor.org/wp-content/uploads/2022/11/TI-Attachment-C.pdf} for details.} Polling companies that are members of the Transparency Initiative outperform those who are not \citep{TIpolls}, suggesting that a firm's willingness to disclose its data collection methods is a proxy for the quality of its estimates. The U.S. federal statistical agencies recently commissioned an expert report on transparency, in an effort to increase trust in federal data \citep{cnstat_transparency}.

Several researchers have similarly called for transparency when releasing benchmark data sets or models \citep{bender-friedman-2018-data, meg2019, Hutchinsonetal2020, Gebru2021, chmielinski2022dataset}. We join these calls and recommend releasing the labeling instructions or guidelines including examples and test questions, the wording of the prompts, information about the labelers, and whether social scientists or domain experts were involved in labeling or consulted on the labeling process. \citet{prabhakaran-etal-2021-releasing, Geiger2020, ulmer-etal-2022-experimental, baan_plank_2024} have also called for better documentation of the label collection process. We commend \citet{nie2020learn, ouyang2022training, glaese2022improving, bai2022training, stiennon2022learning} as particularly good examples of transparency in label collection, with some including screenshots of the label collection instrument. 

Without detailed documentation of data collection methods, researchers will not be able to test many of the hypotheses given above, such as those about wording, task order, ``don't know'' options, and the impact of labeler characteristics. We also suspect that lack of documentation explains difficulties replicating benchmark data sets \citep{recht2019imagenet}. 

\section{Outlook}
Collecting data from labelers is more difficult than the AI/ML literature has recognized. Ambiguous and opinion tasks are particularly challenging, because labelers' responses can be shaped by wording, order, and other context effects as well as by the characteristics of the labeler. We believe that the type of human feedback needed to align future models will resemble opinion collection in surveys. For this reason, greater cooperation and knowledge sharing between the AI/ML and survey methods fields will be crucial to ensuring that the next generation of even more powerful models is human\--centric, trustworthy, and fair. 

Although some hope to replace human labelers with models\--as\--labelers \citep[see, for example,][]{pangakis2023automated, Gilardi_2023, tornberg2023chatgpt4}, the data collection challenges described in this paper will remain. Because models are trained on data collected from humans, as labelers, they can display many of the response biases described in this paper \citep{HardtSurveys, tjuatja2023llms}. In addition, models trained on data labeled by models exhibit unusual behavior, called model collapse or model autophagy disorder \citep{alemohammad2023selfconsuming, gerstgrasser2024model, peterson2024ai}. Models trained on data labeled by models may also exhibit lower performance than those trained on data labeled by humans \citep[see, for example,][]{plaza-del-arco-etal-2024-wisdom}. For the foreseeable future, we will still want humans to create the labels that teach models how to be accurate, fair and safe. Thus, the most important labeling must still be done by humans and is exactly the type of data that is most challenging to collect. 

We applaud the growing interest in data\--centric AI, which focuses on improving models by improving the data they rely on. We recommend that researchers developing AI models be as careful with how their data are collected as they are with their models. We also suggest reaching out to applied statistics departments and social science research groups which often focus on the science of data and may be willing to do the data work.

As is often the case when working across disciplines, as this paper does, language and cultural differences complicate information sharing. The survey methodology literature tends to focus on detecting and measuring bias and variance and attributing it to a cause (interviewers, question wording, survey mode). AI researchers, on the other hand, often understandably focus on solutions, in the form of algorithms distributed through python packages or github repos. These cultural differences between the fields can leave researchers unable to present at each other's conferences or even have productive conversations. Nevertheless, we hope that this position paper has made the case that theory\--driven social science can help AI researchers develop insights and tools that improve the quality of their data and their models.\footnote{We acknowledge that practitioners in AI and ML may be more aware of findings from survey research than the literature suggests.} (And solution\--oriented thinking could also make survey data better.) Perhaps future joint workshops could help the two fields share insights.



\section*{Impact Statement}
This position paper uses results from the field of survey methodology to derive hypotheses about the drivers of data quality in training data collection. We propose several areas for future research and provide concrete ideas to improve the quality of labels collected to train, fine-tune, reinforce, and evaluate AI and ML models. By improving the labels, these ideas will have positive effects on model performance and alignment.

\section*{Acknowledgments}
In writing this paper, we benefited from discussions with Jacob Beck and Bolei Ma of LMU and Rob Chew of RTI International. Fiona Draxler provided valuable comments on an early draft. We would also like to thank the anonymous reviewers for their feedback. BP is supported by the European Research Council (ERC) grant agreements No. 101043235. This research is funded in part by the Bavarian Research Institute for Digital Transformation (bidt) project KLIMA-MEMES (BP) and GREEN DIA (FK), and the German Federal Ministry of Education and Research (BMBF) project KODAQS (16DKZ2019C). The authors are responsible for the content of this publication.

\bibliography{main}
\bibliographystyle{icml2024}


\end{document}